\documentstyle[12pt]{article}
\begin{document}

\begin{center}
\bf{ Zeroth WKB Approximation in Quantum Mechanics }\\
\vspace{3mm}
\rm{ M. N. Sergeenko }\\
\vspace{2mm}
\it{The National Academy of Sciences of Belarus,
    Institute of Physics \\ Minsk 220072, Belarus,\\
    Homel State University, Homel 246699, Belarus \ and \\
    Department of Physics, University of Illinois at Chicago,
                 Illinois 60607, USA }
\end{center}

\begin{abstract}
Solution of the Schr\"odinger's equation in the zero order WKB 
approximation is analyzed. We observe and investigate several 
remarkable features of the WKB$_0$ method. Solution in the whole 
region is built with the help of simple connection formulas we derive 
from basic requirements of continuity and finiteness for the wave 
function in quantum mechanics. We show that, for conservative quantum 
systems, not only total energy, but also momentum is the constant of 
motion. We derive the quantization conditions for two and more 
turning point problems. Exact energy eigenvalues for solvable and 
some ``insoluble'' potentials are obtained. The eigenfunctions have 
the form of a standing wave, $A_n\cos(k_nx+\delta_n)$, and are the 
asymptote of the exact solution.  
\end{abstract}



\section{\label{Intro} Introduction}

The wave equation in quantum mechanics can be derived with the help 
of the Bohr's correspondence principle. This fundamental principle 
has been used at the stage of creation of quantum theory. It is used 
to establish correspondence between classical functions and operators 
of quantum mechanics, and to derive the apparent form of the 
operators \cite{Bohm,Shf}. Moreover, the correspondence principle 
points out the way to a simplest asymptotic solution of the 
Schr\"odinger's equation.

The correspondence principle states that the laws of quantum physics 
must be so chosen that in the classical limit, where many quanta are 
involved, the quantum laws lead to the classical equations as an 
average. In this way, in Ref. \cite{SeSe} this principle has been 
used to derive the nonrelativistic semiclassical wave equation 
appropriate in the quasiclassical region and, in Ref. \cite{SeRe}, 
the corresponding relativistic wave equation. It was shown that 
application of the standard WKB method to the semiclassical wave 
equation allows to solve the so called ``insoluble'' problems 
\cite{SeRe,SeCl,SeKr} and results in the existence quantum 
fluctuations of the angular momentum, which contribute to the energy 
of the ground state \cite{SeF}.

There is a mathematical realization of the correspondence principle 
known as the quasiclassical approximation which is widely used mainly 
as the Wentzel-Kramers-Brillouin (WKB) method \cite{Fro,Mertz,Len,Tr} 
applicable in the case when the de Broglie wavelength, $\lambda = 
h/p$ ($h=2\pi\hbar$), is changing slowly. This method is usually used 
as a tool to obtain the approximate solution of the one-dimensional 
Schr\"odinger's equation in the quasiclassical region at large values 
of quantum numbers.

In this work, we analyze the classical limit of the Schr\"odinger's 
wave mechanics with the help of the correspondence principle. We 
investigate solution of the wave equation in the {\em zero-order} WKB 
approximation (WKB$_0$) and observe several remarkable features and 
important advantages of the zeroth approximation with reference to 
the commonly used first-order WKB approximation (WKB$_1$). These 
advantages are: 1) the WKB$_0$ wave function ({\em wf}), $\psi_0(x)$, 
has no divergencies at the turning points; 2) the WKB$_0$ 
approximation allows a considerably simpler (than the standard 
approach based on linear approximation) derivation of the connection 
formulas and quantization condition; 3) the WKB$_0$ method allows to 
solve the wave equation for many-turning-point problems; 4) for 
stationary states, not only energy, but also momentum is the constant 
of motion; and, finally, 5) the most important advantage is that the 
WKB$_0$ approach reproduces the exact energy spectra for all known 
solvable potentials and ``insoluble'' potentials as well.

Solution of the wave equation in the whole region is built with the 
help of connection formulas we derive in this work, which give the 
main term of the asymptotic series. To demonstrate efficiency of the 
WKB$_0$ approach, we calculate exact energy eigenvalues for solvable 
and some ``insoluble'' spherically symmetric potentials.  For 
discrete spectrum, the WKB$_0$ eigenfunctions are written in terms of 
elementary functions in the form of a standing wave, 
$A_n\cos(k_nx+\delta_n)$, and give the asymptote of the exact 
solution.

\section{\label{ZeroWKB} The zero order WKB solution}

It is well known that the exact eigenvalues can be defined with the 
help of the asymptotic solution, i.e. the exact solution and its 
asymptote correspond to the same exact eigenvalues of the problem 
under consideration. The asymptotic solution in quantum mechanics can 
be obtained by the WKB method. This means that the quasiclassical 
method can be considered as a tool to reproduce the exact energy 
spectrum. This is true for some potentials \cite{Dunh,Krei,Kre,Ros} 
and has proven for many others with the help of specially developed 
techniques, improvements, or modifications of the quasiclassical 
method on the real axis and in the complex plane 
\cite{Fro,Dunh,Art,Bru,Lang,SeQ}.

The main problem concerning the applicability of the WKB 
approximation is the problem of exactness of the method. Proofs of 
varying degrees of rigor have been advanced that demonstrate the 
exactness of the standard WKB quantization condition 
\cite{Fro,Mertz,Len,Tr,Dunh,Krei,Kre,Ros,Art,Bru,Lang,SeQ}.  The 
question of exactness of the WKB approximation is usually reduced to 
the estimation of the high-order correction terms. The earliest 
development of the WKB method for obtaining the high-order 
corrections has been considered in Ref. \cite{Dunh}. Then, in 
\cite{Krei}, the radial generalization of Dunham's one-dimensional 
WKB quantization condition was derived with the help of the Langer 
transformation \cite{Lang}. It was shown that the second- and 
third-order integrals identically vanish for the hydrogen atom and 
the three-dimensional harmonic oscillator \cite{Krei}. As we show 
below, the first order term also does not affect the exactness of the 
quasiclassical quantization condition.

Consider the static Schr\"odinger's equation with the arbitrary 
potential $V(x)$ in one dimension,

\begin{equation}  \label{SchE}
\left[\frac 1{2m}\left(-i\hbar\frac d{dx}\right)^2+V(x)\right]\psi(x)=
E\psi(x).
\end{equation}
Let us write the equation (\ref{SchE}) in the form,

\begin{equation} \label{SchP}
\left(-i\hbar\frac d{dx}\right)^2\psi(x)=\left[P^2-U(x)\right]\psi(x)~,
\end{equation}
where $P^2=2mE$ and $U(x)=2mV(x)$. Equations (\ref{SchE}) and 
(\ref{SchP}) are equivalent, but the physical sense of their 
solutions is different. The Schr\"odinger's equation (\ref{SchE}) is 
the energy eigenvalue problem, and equation (\ref{SchP}) is the 
eigenvalue problem for the squared momentum $P^2$. Equation 
(\ref{SchP}) has two solutions for each eigenstate (as it should be 
for the second-order differential equation) with the corresponding 
eigenvalues $P_n^{(1,2)}=\pm\sqrt{2mE_n}$, i.e. eigenvalues for these 
two problems are connected via the equality $P_n^2=2mE_n$ for a free 
particle.

In the conventional WKB method, solution of Eq. (\ref{SchP}) is 
searched for in the form

\begin{equation} \label{psix}
\psi(t,x) = A\exp\left[\frac i\hbar S(t,x)\right],
\end{equation}
where $A$ is the arbitrary constant and $S(t,x)$ is the complex 
function.  Substituting (\ref{psix}) into Eq. (\ref{SchP}) one finds 
that $S(t,x)$ satisfies the equation

\begin{equation} \label{HJeq}
\left(\frac{dS}{dx}\right)^2 -
i\hbar\frac d{dx}\left(\frac{dS}{dx}\right) = P^2-U(x).
\end{equation}
The Hamilton-Jacobi equation in quantum mechanics (\ref{HJeq}) is 
equivalent to the wave equation (\ref{SchP}). In the limit 
$\hbar\rightarrow 0$ Eq. (\ref{HJeq}) reduces to the classical 
Hamilton-Jacobi equation.

In the WKB approximation one expands $S(t,x)$ in power series of 
$\hbar$, $S=S_0+$ $\hbar S_1+$ $\hbar^2S_2+\ldots$ and usually keeps 
two first terms of the expansion, i.e. $S_0(t,x)$ and $S_1(t,x)$ (the 
WKB$_1$ approximation). This gives, after simple calculations, 
$(dS_0/dx)^2= P^2-U(x)$, and $S_1(t,x)= 
i\ln\sqrt{\left|p(x)\right|}$, where $p(x)=dS_0/dx$. For conservative 
systems, the classical action $S_0(t,x)$ can be written as $S_0(t,x)= 
-Et+W(x)$, where $W(x)$ is the Hamilton's function (reduced classical 
action). Hence, one gets

\begin{equation} \label{HJ}
\left(\frac{dW}{dx}\right)^2 = P^2-U(x).
\end{equation}
Here $dW/dx\equiv p(x)=\pm\sqrt{P^2-U(x)}$ is the generalized 
momentum and the WKB$_1$ {\em wf} is

\begin{equation}  \label{fS1}
\psi_{WKB}^{(1)}(x) = \frac 1{\sqrt{\left|p(x)\right|}}
\left[C_1e^{i\phi(x)} + C_2e^{-i\phi(x)} \right],
\end{equation}
where $\phi(x)$ is dimensionless phase variable,

\begin{equation}  \label{phi}
\phi(x) = \frac 1\hbar W(x)\equiv
\frac 1\hbar\int^x\sqrt{P^2-U(x)}dx~.
\end{equation}

There are several disadvantages of the WKB$_1$ approximation. The 
most essential of them is that the WKB$_1$ {\em wf} (\ref{fS1}) 
diverges at the classical turning points given by $p(x)=0$, where the 
condition for its applicability breaks down. Although this divergence 
is understandable in the classical limit, since a classical particle 
has zero speed at the turning points, it is certainly not present in 
a quantum mechanical treatment. Since $\psi_{WKB}^{(1)}(x)$ is 
singular, tricky matching the {\em wf} $\psi_{WKB}^{(1)}(x)$ and its 
first derivative at the classical turning points have been developed 
\cite{Lang} that yields the famous connection formulas and well known 
WKB quantization condition. The exact mathematical conditions under 
which the connection formulas can be proved rigorously to apply are 
fairly complex. Other difficulties arising in the WKB$_1$ 
approximation have been considered elsewhere \cite{SeSe}.

The connection formulas and quantization condition can be obtained 
much simpler in the WKB$_0$ approximation \cite{SeA,SeSy}, i.e. in 
the classical limit $\hbar\rightarrow 0$. In this case $S(t,x)\simeq 
S_0(t,x)$ and the general solution of the wave equation (\ref{SchP}) 
for the potential $V(x)$ can be written in terms of the dimensionless 
phase variable $\phi(x)$,

\begin{equation}  \label{psi0}
\psi_{WKB}^{(0)}(x)\equiv\psi_0[\phi(x)] =
C_1e^{i\phi(x)} + C_2e^{-i\phi(x)}~,
\end{equation}
where $C_1$ and $C_2$ are taken constant in the classical limit.

The WKB$_0$ solution (\ref{psi0}) has several important advantages 
with reference to the commonly used WKB$_1$ {\em wf} (\ref{fS1}). The 
most important is that the solution (\ref{psi0}) has no divergence at 
turning points. In the phase space, the function (\ref{psi0}) has the 
form of superposition of two plane waves; it has also such a form in 
the configuration space (see below). Note, that the WKB$_0$ 
approximation corresponds to the main term of the asymptotic series 
in the theory of the second-order differential equations.

\section{\label{ConnecForQuan} Connection formulas. Quantization}

Basic requirements for the {\em wf} in quantum mechanics are 
continuity and finiteness in the whole region. The famous connection 
formulas \cite{Lang} have been derived to connect the WKB solution in 
the classical region with the one in non-classical region. These 
formulas are based on a linear approximation to the potential in the 
regions around the classical turning points where the WKB$_1$ {\em 
wf} diverges; the exact solution involves Bessel's functions of order 
$1/3$. Even though the connection formulas themselves turn out to be 
simple enough, their derivation is quite tedious.

One of the most important problems to which the connection formulas 
apply is that of penetration of a potential barrier. Interaction with 
the potential barrier can be of two types - reflection and 
transmission. There are several problems in which the conventional 
WKB$_1$ formulas break down \cite{Bohm}. One kind of problems occurs 
near the top of a barrier, where the slope of the potential is small.  
As a result, the straight-line approximation to the potential breaks 
down, and the connection formulas must be altered. The second type of 
problems arises when the potential changes too rapidly, for example, 
in the case of a square well potential. The connection formulas also 
break down for some other type of potentials \cite{Bohm}.

Consider the WKB$_0$ {\em wf} (\ref{psi0}) for the two-turning-poing 
(2TP) problem. To build the physical solution in the whole region we 
need to connect the oscillating solution in classically allowed 
region with the exponentially decaying solutions in classically 
inaccessible regions. For the 2TP problem, the whole interval 
($-\infty,\infty$) is divided by the turning points $x_1$ and $x_2$ 
into three regions, $(-\infty,x_1]$ (region I), $[x_1,x_2]$ (region 
II), and $[x_2,\infty)$ (region III). In the classically allowed 
region II, right from the turning point $x_1$, the real part of 
solution (\ref{psi0}) can be written in the form

\begin{equation}  \label{cs1}
\psi_0^{II}(\phi) = C_1\cos(\phi -\phi_1 + \delta_1),
\end{equation}
and left from the turning point $x_2$,

\begin{equation}  \label{cs2}
\psi_0^{II}(\phi) = C_2\cos(\phi -\phi_2 + \delta_2),
\end{equation}
where $\phi_1=\phi(x_1)$, $\phi_2=\phi(x_2)$, and $\delta_{1,2}$, are 
the phase shifts at the turning points $x_1$ and $x_2$, respectively.  
In the classically inaccessible regions I and III, to guaranty 
finiteness of the {\em wf} at $x\rightarrow \pm\infty$, we choose the 
exponentially decaying solutions, i.e.,

\begin{equation}   \label{ex1}
\psi_0^I(\phi) = Ae^{\phi -\phi_1},
\end{equation}
left from the turning point $x_1$, and

\begin{equation}  \label{ex2}
\psi_0^{III}(\phi) = Be^{-\phi +\phi_2},
\end{equation}
right from the turning point $x_2$.

To construct solution in the whole region we need to connect 
functions $\psi_0^I(\phi)$, $\psi_0^{II}(\phi)$, and 
$\psi_0^{III}(\phi)$ at the turning points $x_1$ and $x_2$. Matching 
the functions $\psi_0^I(\phi)$ and $\psi_0^{II}(\phi)$ and their 
first derivatives at the turning point $x=x_1$ gives

\begin{equation}
\left\{
\begin{array}{l}  \label{ed1}
C_1\cos\delta_1 = A,\\
C_1\sin\delta_1 = -A,
\end{array}
\right.
\end{equation}
and for $\psi_0^{II}(\phi)$ and $\psi_0^{III}(\phi)$ at the turning 
point $x=x_2$ we have

\begin{equation}
\left\{
\begin{array}{l}  \label{ed2}
C_2\cos\delta_2 = B,\\
C_2\sin\delta_2 = B,
\end{array}
\right.
\end{equation}
which yields

\begin{equation}
\left\{
\begin{array}{l}  \label{d12}
\tan\delta_1 =-1,\\
\tan\delta_2 = 1,
\end{array}
\right.
\end{equation}
and

\begin{equation}
\left\{
\begin{array}{l}  \label{d12C}
C_1=\sqrt 2A,\\
C_2=\sqrt 2B.
\end{array}
\right.
\end{equation}

The connection formulas (\ref{d12}) and (\ref{d12C}) supply the 
continuous transition of the oscillating solutions (\ref{cs1}) and 
(\ref{cs2}) into exponential solutions (\ref{ex1}) and (\ref{ex2}) at 
the turning points $x_1$ and $x_2$.

The function $\psi_0^{II}(\phi)$ [given by Eqs. (\ref{cs1}) and 
(\ref{cs2})] should match itself at each point of the interval 
$[x_1,x_2]$. Putting $x=x_2$ we have, from Eqs. (\ref{cs1}) and 
(\ref{cs2}),

\begin{equation}  \label{cs12}
C_1\cos(\phi_2-\phi_1 + \delta_1) = C_2\cos\delta_2.
\end{equation}
This equation is valid if

\begin{equation} \label{eq}
\phi_2 - \phi_1 + \delta_1 = \delta_2 + \pi n ,~~
n = 0,1,2,\ldots
\end{equation}
and $C_2=(-1)^nC_1$.
Equation (\ref{eq}) is the condition of the existence of continuous 
finite solution in the whole region. This condition being, at the 
same time, the quantization condition. Taking into account the 
notation (\ref{phi}), we have, from Eq. (\ref{eq}),

\begin{equation}  \label{qc2}
\int_{x_1}^{x_2}\sqrt{P^2-U(x)}dx = \pi\hbar\left(n +\frac 12\right).
\end{equation}

Condition (\ref{qc2}) solves the 2TP eigenvalue problem given by Eq.  
(\ref{SchP}). We see that the WKB$_0$ approximation results in the 
quantization of the reduced classical action, $W(x)$, and coincides 
with the well known WKB quantizaton condition. It is important to 
underline that the first order term does {\em not} affect the 
exactness of the quantization condition. In our case, the condition 
(\ref{qc2}) is obtained by product from general requirements of 
continuity and finiteness to the wave function in the whole region, 
i.e. from the requirements of a smooth transition of the oscillating 
solution given by Eqs. (\ref{cs1}) and (\ref{cs2}) to the 
exponentially decaying solutions (\ref{ex1}) and (\ref{ex2}) in the 
classically inaccessible regions. Quantization condition (\ref{qc2}) 
reproduces the exact eigenvalues for {\em all} known 2TP problems 
(see Refs. \cite{SeSe,SeRe,SeQ}).

The derivation given above is evidently much simpler than the usual 
text book approach for deriving connection formulas and quantization 
condition. In the general case of three or more turning points, phase 
shifts $\delta_1$ and $\delta_2$ will be different \cite{SeSu}.

Combining the above results, we can write the finite continuous 
WKB$_0$ solution in the whole region ($\delta_2=-\delta_1=\pi /4$),

\begin{equation}
\psi_0[\phi(x)] = C\left\{
\begin{array}{lc}  \label{osol}
\frac 1{\sqrt 2}e^{\phi(x) -\phi_1}, & x<x_1,\\
\cos[\phi(x) -\phi_1 -\frac\pi 4], & x_1\le x\le x_2,\\
\frac{(-1)^n}{\sqrt 2}e^{-\phi(x) +\phi_2}, & x>x_2.
\end{array}
\right.
\end{equation}
Oscillating part of solution (\ref{osol}) corresponds to the main 
term of the asymptotic series in theory of the second-order 
differential equations. In quantum mechanics, the oscillating part of 
Eq. (\ref{osol}) gives the asymptote of the exact solution of the 
Schr\"odinger's equation (see below).

\section{\label{MomConsrv} Momentum conservation for discrete spectrum}

It is well known that, for discrete spectrum, the asymptote of Eq.  
(\ref{SchP}) has the form $\psi_n(x)= A_n\cos(k_nx+\delta_n)$, where 
$A_n$, $k_n=P_n/\hbar$, and $\delta_n$ are constant values; here 
$k_n$ is the wave number and $P_n$ is the momentum eigenvalue. The 
asymptote has the form of a standing wave and can be written as a 
superposition of two plane waves, $\exp(ik_nx)$ and $\exp(-ik_nx)$, 
each of which describes free motion. Therefore, for the discrete 
spectrum, the asymptote of Eq. (\ref{SchP}) describes free motion of 
a particle-wave in the enclosure and the eigenmomentum $P_n=\hbar 
k_n$ is the constant of motion.

The quasiclassical approximation is the asymptotic method, which 
takes advantage of the fact that the wave length is changing slowly, 
by assuming that the {\em wf} is not changed much from the form it 
would take if the potential $V$ was constant. The WKB$_0$ solution 
(\ref{psi0}) is justified in the classical limit $\hbar\rightarrow 
0$, which corresponds to the short-wave asymptote of the 
Schr\"odinger's equation (\ref{SchP}). Our basic requirements to the 
solution (\ref{psi0}) are: 1) $C_1$ and $C_2$ are constant values and 
2) the generalized momentum $p(x)=dW/dx$ is the adiabatically slowly 
changing function, i.e.,

\begin{equation}  \label{cotr1}
\frac{dW}{dx}\simeq {\rm const}.
\end{equation}
These are the same assumptions which have been originally used by 
Schr\"odinger to derive the wave equation from the optical-mechanical 
analogy (see, for instance, Ref. \cite{Bohm}).

In this analogy, the most important aspect is the relation to the 
limiting case of wave optics, i.e. geometrical optics. The limiting 
case of geometrical optics is analogous to the classical limit of 
quantum mechanics: the amplitude of the ({\em wf}) is a constant and 
momentum is the adiabatically slowly changing function. Note, the 
constraint (\ref{cotr1}) supplies the Hermiticity of the operator 
$\hat p^2=[-i\hbar(d/dx)]^2$ in Eq. (\ref{SchP}) \cite{SeSe}.

Consider the WKB$_0$ eigenfunction (\ref{osol}) in the classically 
allowed region given by $p(x)>0$. Show that the oscillating part of 
the {\em wf} (\ref{osol}) is in agreement with the asymptote of the 
corresponding exact solution of Eq. (\ref{SchP}) and can be written 
in the form of a standing wave. In equations (\ref{SchE}) and 
(\ref{SchP}), the total energy $E$ and the momentum $P$ are connected 
by means of the equality $P^2=2mE$ for a free particle. Integrating 
(\ref{cotr1}), we obtain, for the action $W(x)$,

\begin{equation}
W(x) \simeq Px + {\rm const},  \label{act0}
\end{equation}
where $P=\sqrt{2mE}$ is the total momentum. Solution of Eq. 
(\ref{SchE}) [or (\ref{SchP})] results in the energy quantization, 
i.e. $E=E_n$.  Therefore, $W(x)\simeq P_nx+{\rm const}$, where $P_n 
=\hbar k_n= \sqrt{2mE_n}$. Underline, this form of the action is 
valid only for allowed motions in quantum mechanics, i.e. for the 
energies $E=E_n$, which can be found with the help of the 
quantization condition (\ref{qc2}).

Thus, the oscillating part of the {\em wf} (\ref{osol}) takes the
form

\begin{equation}  \label{fn0}
\psi_0^{II}(x)= C_n\cos\left(\frac 1\hbar P_nx +\frac\pi 2 n\right).
\end{equation}
The normalization coefficient,

\begin{equation}  \label{Cn}
C_n = \sqrt{\frac{2P_n}{\pi\hbar(n+\frac 12)+\hbar}},
\end{equation}
is calculated from the normalization condition

\begin{equation}
\int_{-\infty}^\infty\left|\psi_0(x)\right|^2dx=1.
\end{equation}
In Eq. (\ref{fn0}), we have took into account the fact that, in the 
stationary states, the phase-space integral (\ref{phi}) at the TP 
$x_1$ and $x_2$ is $\phi_1=-\pi(n+1/2)/2$ and $\phi_2=\pi(n+1/2)/2$, 
respectively, so that $\phi_2- \phi_1= \pi(n+1/2)$ \cite{SeSe,SeQ}.  
The phase integrals $\phi_1$ and $\phi_2$ depend on quantum number 
and do not depend on the form of the potential. This form of $\phi_1$ 
and $\phi_2$ guaranties that the eigenfunctions are necessarily 
either symmetrical ($n=0,2,4,\ldots$) or antisymmetrical 
($n=1,3,5,\ldots$).

Solution (\ref{fn0}) has the form of a standing wave, which can be 
written as the superposition of two plane waves. The fact that, for 
stationary states, the oscillating part of solution (\ref{fn0}) can 
be written in such form means that (for conservative 
quantum-mechanical systems) the particle momentum is the {\em 
constant of motion} and the corresponding coordinate is cyclical. 
Therefore, solution (\ref{fn0}) describes free motion of a 
particle-wave in the enclosure, where the enclosure is the 
interaction potential. This means that interaction of the 
particle-wave with the potential reduces to reflection of the wave by 
the walls of the potential.

As other integrals of motion in quantum mechanics, the total momentum 
$P$ can take only discrete values, $P=P_n$, corresponding to the 
discrete values of the classical action $W(x)$, which can be found 
from the quantization condition (\ref{qc2}). Note that, for a 
particle moving in the Coulomb potential, the energy eigenvalues can 
be written in the form of the total energy for a free particle, i.e. 
$E_n=P_n^2/2m$, where $P_n= i\alpha m(n\hbar)^{-1}$ is the momentum 
eigenvalue and $n$ is the principal quantum number. It is easy to 
show that the energy eigenvalues for other solvable potentials also 
can be written in such form.

There are the following physical grounds to the above discussions.  
According to classical theory, accelerated electrons radiate energy 
at a rate equal to $\frac 23e^2c^{-3}$\"x$^2$. But electrons in 
atomic orbits are {\em always accelerated (?)} and should, therefore, 
lose energy continuously until they fall into the nucleus. Actually, 
it is known that this never happens. To resolve this problem, Bohr 
postulated that 1) the electrons are in stationary states, i.e. do 
not radiate, {\em despite their acceleration} and 2) electrons can 
take discontinuous transitions from one allowed orbit to another.  
However, the postulate is not an answer to the question: why the 
accelerated electrons do not radiate?

Acceleration means change in momentum. In quantum mechanics, the 
change in momentum (which is associated with the de Broglie wave 
$\lambda =h/p$, $h=2\pi\hbar$) appears as radiation of frequency 
$\nu=(E-E^\prime)/h$ that means change of the stationary state. As we 
have shown above, in the stationary states, particles are {\em not} 
accelerated, because the momentum is the constant of motion. This 
means that, in the stationary states, the electrons  move as free 
particles, particles-waves in enclosures.

The well known quasiclassical condition,

\begin{equation}  \label{qcond}
\frac\hbar{p^2}\left|\frac{dp}{dx}\right|\ll 1~,
\end{equation}
defines the region of applicability of the WKB$_1$ approximation.  
Conventional treatment of this condition implies that the momentum 
$p(x)$ is large enough, i.e. quantum number $n$ take large values.  
At the same time, the WKB method yields the exact eigenvalues for 
{\em all} values of $n$, including $n=0$ (see Ref. 
\cite{SeSe,SeRe,SeQ}).  This means that the condition (\ref{qcond}) 
can be treated differently:  it supplies the hermiticity of the 
squared momentum operator, $\hat p^2$, that implies $dp/dx\simeq 0$, 
and does not imply that the momentum $p(x)$ takes large values 
\cite{SeSe}.

\section{\label{MultiTP} The multi-turning-point problems}

The WKB method is usually used to solve one-dimensional two turning 
point problems.  Within the framework of the WKB method the solvable 
potentials mean those potentials for which the eigenvalue problem has 
two turning points. However the WKB$_0$ solution (\ref{psi0}) is 
general for all types of problems. The zeroth WKB approximation 
allows a considerably simpler derivation of the connection formulas 
and the corresponding quantization condition not only for two but, 
also, for many-turning-point problems ($\mu$TP, $\mu>2$). This is the 
class of the so-called ``insoluble'' problems, which cannot be solved 
by standard methods.

The quasiclassical formulas can be written both on the real axis and 
in the complex plane. Most general form of the WKB solution and 
quantization condition can be written in the complex plane. In the 
complex plane, the 2TP problem has one cut between turning points 
$x_1$ and $x_2$, and the phase-space integral (\ref{qc2}) can be 
written as the contour integral about the cut. The $\mu$TP problems 
contain (in general case) bound state regions and the potential 
barriers, i.e. several cuts. The corresponding contour $C$ should 
enclose all cuts. The WKB$_0$ {\em wf} in the whole region can be 
built similarly to the 2TP one considered above.

Let the problem has $\nu$ finite cuts, i.e. $\mu=2\nu$ turning 
points, and the potential is infinite between cuts\footnote{In case 
where the potential is finite in the whole region, the quantization 
condition will be more complicate \cite{SeSu}.}. Then the integral 
about the contour $C$ can be written as a sum of contour integrals 
about each of the cut,

\begin{equation}   \label{oink}
\oint_C\sqrt{P^2-U(x)}dx = \sum_{i=1}^{\nu}\oint_{C_i}\sqrt{P^2-U(x)}dx,
\end{equation}
where each term of the sum is the 2TP problem,

\begin{equation}  \label{qci}
\oint_{C_i}\sqrt{P^2-U(x)}dx = 2\pi\hbar\left(n_i +\frac 12\right).
\end{equation}
Therefore, the $\mu$TP quantization condition is \cite{SeSe,SeQ}

\begin{equation}  \label{genCi}
\oint_C\sqrt{P^2-U(z)}dz =
 2\pi\hbar\sum_{i=1}^{\nu}\left(n +\frac 12\right)_i.
\end{equation}

The condition (\ref{genCi}) is in agreement with the Maslov's theory 
\cite{MaslFed}. This means that the right-hand side of the equation 
(\ref{oink}) can be written as

\begin{equation}
2\pi\hbar\sum_{i=1}^{\nu}\left(n +\frac 12\right)_i =
2\pi\hbar\left(N +\frac\mu 4\right).  \label{qmas}
\end{equation}
where $N=\sum_{i=1}^{\nu}n_i$ is the total number of zeroes of the 
{\em wf} on the $\nu$ cuts. In this interpretation, the number of 
turning points $\mu$ is the Maslov's index, i.e. number of 
reflections of the {\em wf} on the walls of the potential. Thus the 
condition (\ref{genCi}) takes the form

\begin{equation}  \label{genC}
\oint_C\sqrt{P^2-U(z)}dz = 2\pi\hbar\left(N +\frac\mu 4\right).
\end{equation}

\section{\label{MultiDim} Multi-dimensional problems}

In most practical applications we deal with the multi-dimensional 
problems.  For known solvable spherically symmetric potentials, the 
conventional WKB method does not reproduce the exact energy levels 
unless one supplements it with Langer-like correction terms.

To overcome this problem in particular case of the Coulomb potential, 
a special techniques has been developed. In order for the WKB$_1$ 
approximation to give the exact eigenvalues, the quantity $l(l+1)$ in 
the radial equation must be replaced by $(l+1/2)^2$ \cite{Lang}.  The 
reason for this modification (for the special case of the Coulomb 
potential) was pointed out by Langer ($1937$) \cite{Lang} from the 
Langer transformation $r = e^x$, $U(r) = e^{x/2}X(x)$. However, for 
other spherically symmetric potentials, in order to obtain the 
appropriate Langer-like correction terms, another special 
transformation of the {\em wf} and its arguments is required.

There are several other related problems in the semiclassical 
consideration of the radial Schr\"odinger equation. (i) The WKB 
solution of the radial equation is irregular at $r\rightarrow 0$, 
i.e. $R^{WKB}(r)\propto r^\lambda /\sqrt r$, $\lambda = 
\sqrt{l(l+1)}$, whereas the exact solution in this limit is 
$R(r)\propto r^l$. (ii) The radial equation has no the centrifugal 
term when $l=0$, i.e. the radial problem has only one turning point 
and one can not use the WKB quantization condition derived for 
two-turning-point problems. However, solving the equation for $l>0$ 
by known exact methods one obtains energy eigenvalues for all $l$. 
(iii) The WKB solution of the angular equation has analogous to the 
radial one, incorrect behavior at $\theta\rightarrow 0$: 
$\Theta^{WKB}(\theta) \propto\theta^\mu$, $\mu^2 = m^2 -\hbar^2/4$, 
while the exact regular solution in this limit is 
$\Theta_l^m(\theta)$ $\propto \theta^{|m|}$. Angular eigenfunction 
$Y_{00}(\theta,\varphi) = {\rm const}$, i.e. no nontrivial solution 
exists.

Practical use shows the standard leading-order WKB approximation {\em 
always} reproduces the exact spectrum for the solvable spherically 
symmetric potentials $V(r)$ if the centrifugal term in the radial 
Schr\"odinger's equation has the form $(l+1/2)^2\hbar^2/r^2$.  As was 
shown in Ref. \cite{SeQ} the centrifugal term of such a form can be 
obtained from the WKB solution of the angular Schr\"odinger's 
equation if separation of the original three-dimensional equation has 
performed with the help of the correspondence principle.

In Refs. \cite{SeSe,SeQ} we have fulfilled the quasiclassical 
analysis of the three-dimensional Schr\"odinger's equation and 
suggested the quasiclassical approach for multi-dimensional problems. 
In this approach the original wave equation is reduced to the form of 
the classical Hamilton-Jacobi equation without first derivatives. It 
was shown that, in the quasiclassical region the wave equation can be 
written in the form \cite{SeSe,SeQ},

\begin{equation}  \label{RedSh}
(-i\hbar )^2\left[\frac{\partial ^2}{\partial r^2}+\frac
1{r^2}\frac{ \partial ^2}{\partial \theta ^2}+\frac
1{r^2\sin^2\theta }\frac{\partial ^2}{ \partial \varphi
^2}\right] \tilde \psi (\vec r) =  \nonumber  \\
2m\left[ E-V(r)\right]\tilde \psi (\vec r).
\end{equation}

Separation of this equation is performed with the help of the 
correspondence principle between classical and quantum-mechanical 
quantities. As a result of the separation, we have obtained a system 
of reduced second-order differential equations,

\begin{equation} \label{RadEq}
\left(-i\hbar\frac{d}{dr}\right)^2\tilde R(r) = \left[2m
\left(E-V(r)\right)- \frac{\vec M^2}{r^2}\right]\tilde R(r),
\end{equation}

\begin{equation} \label{AngEq}
\left(-i\hbar\frac{d}{d\theta}\right)^2\tilde{\Theta}(\theta) =
\left[ \vec M^2-\frac{M_z^2}{\sin^2\theta}\right] \tilde{\Theta}
(\theta),
\end{equation}

\begin{equation} \label{FiEq}
\left(-i\hbar\frac d{d\varphi}\right)^2\tilde\Phi(\varphi
)=M_z^2\tilde\Phi(\varphi),
\end{equation}
where $\vec M^2$, $M_z^2$ are the constants of separation and,
at the same time, constants of motion.

Each of Eqs. (\ref{RadEq})-(\ref{FiEq}) has the general WKB$_0$ 
solution of the form (\ref{psi0}). Application of the quantization 
condition (\ref{qc2}) to the angular equation (\ref{AngEq}) results 
in the squared angular momentum eigenvalues \cite{SeSe,SeQ},

\begin{equation} \label{M2}
\vec M^2 =\left(l+\frac 12\right)^2\hbar^2.\\
\end{equation}

The same quantization condition appropriate to the radial equation 
(\ref{RadEq}) yields the exact energy spectra for all known solvable 
potentials and ``insoluble'' potentials, as well.

\section{\label{Applic} Some practical applications}

To illustrate efficiency of the WKB$_0$ method consider several 
classic problems. The general solution (\ref{psi0}) is the same for 
all problems in quantum mechanics. In case of multi-dimensional 
problems, we use the corresponding multi-dimensional classical 
action. The oscillating part of the WKB$_0$ {\it wf} (\ref{fn0}) 
gives the asymptote of the exact solution of the Schr\"odinger's 
equation. Most complicate computational part in this approach is 
calculation of the phase-space integral. In case of multi-dimensional 
problems, separation should be performed with the help of the 
correspondence principle between classical and quantum-mechanical 
quantities \cite{SeF,SeQ}.

\subsection{\label{LinHarmOsc} The linear harmonic oscillator
$V(x)=\frac 12 m\omega^2 x^2$}

Consider first the linear harmonic oscillator.  Many problems in 
physics can be reduced to a harmonic oscillator with appropriate 
conditions. The general differential equation for oscillator 
potential can be solved using a technique that is frequently 
exploited in solving quantum mechanics problems. The eigenfunctions 
that are the solutions of the Schr\"odinger equation are the Hermite 
polynomials. In our approach, we directly obtain the asymptote of the 
exact solution and the corresponding well known exact energy 
eigenvalues.

The quantization condition (\ref{qc2}) appropriate to the 
Schr\"odinger equation with the linear oscillator potential is

\begin{equation}  \label{losc}
I=\int_{r_1}^{r_2}\sqrt{2mE - (m\omega x)^2}dx =
\pi\hbar\left(n+\frac 12\right).
\end{equation}
The phase-space integral (\ref{losc}) is calculated in closed form, 
$I=E/\omega$, that results in the exact energy eigenvalues, 
$E=\omega\hbar(n+1/2)$. The WKB$_0$ {\it wf} of the oscillator (the 
asymptote of the problem) is given by Eq. (\ref{fn0}) and has the 
form of the standing wave,

\begin{equation}  \label{wflo}
\psi_0(x) = C_n\cos\left(\frac 1\hbar P_nx + \frac\pi 2 n\right).
\end{equation}
The normalization constant is calculated according to Eq. (\ref{Cn}) 
with $P_n=\sqrt{2mE_n}$.

\subsection{\label{CoulPot} The Coulomb problem $V(r)= -\alpha/r$}

The Coulomb potential is a classic example of the exactly solvable 
problems in quantum mechanics. For this 2TP problem, the standard WKB 
method does not reproduce the exact energy levels unless one 
supplements it with Langer-like correction terms. In our approach, 
separation of the original 3D Schr\"odinger's equation with the help 
of the correspondence principle \cite{SeQ} results in the centrifugal 
term $\vec M^2/r^2$ with $\vec M^2$ given by Eq. (\ref{M2}), which 
does not require any Langer type corrections. The quantization 
condition (\ref{qc2}), for the Coulomb problem, is

\begin{equation}   \label{Icou}
I =\oint_C\sqrt{2mE +\frac{2m\alpha}r - \frac{\vec M^2}{r^2}}dr =
2\pi\hbar\left(n_r+\frac 12\right),
\end{equation}
where the integral is taken about a contour $C$ inclosing the turning 
points $r_1$ and $r_2$. To calculate the integral (\ref{Icou}) we use 
the method of stereographic projection. To do this, we should exclude 
the singularities outside the contour $C$, i.e. at $r=0$ and 
$\infty$.  Excluding these infinities we have, for the integral 
(\ref{Icou}), $I = I_0 + I_{\infty}$, where $I_0= -2\pi M$ and 
$I_{\infty}= 2\pi i\alpha m/ \sqrt{2mE}$. The sequential simple 
calculations result in the exact energy spectrum

\begin{equation} \label{Ecou}
 E_n = -\frac{\alpha^2m}{2[(n_r+\frac 12)\hbar + M]^2}.
\end{equation}
Note, for the energy of zeroth oscillations we have, from Eq.  
(\ref{Ecou}), $E_0 = -\alpha^2 m/[2(\hbar/2 + M_0)^2]$, that 
apparently shows the contribution of the quantum fluctuations of the 
angular momentum, $M_0=\hbar/2$, into the energy of the ground state 
$E_0$ \cite{SeF}. The radial WKB$_0$ eigenfunctions, $R_n^{(0)}(r)$, 
inside the classical region $[r_1,r_2]$ are written according to Eq. 
(\ref{fn0}) and give the asymptote of the problem,

\begin{equation} \label{Rn}
\tilde R_n^{(0)}(r)= C_n\cos\left(\frac 1\hbar P_nr +\frac\pi 2 n_r\right),
\end{equation}
The normalization constant, $C_n$, is calculated with the help of Eq.  
(\ref{Cn}), where $P_n=\sqrt{2m|E_n|}$ and $E_n$ is given by Eq.  
(\ref{Ecou}).

\subsection{\label{IsotrOsc} The isotropic oscillator
$V(r)=\frac 12 m\omega^2r^2$}

The three-dimensional harmonic oscillator is another classic example 
of the exactly solvable problems in quantum mechanics. As in case of 
the Coulomb potential, the WKB method does not reproduce the exact 
energy levels unless one supplements it with Langer-like correction 
terms. This 4TP problem is usually solved with the help of the 
replacement $z=r^2$, which reduces the problem to the 2TP one.

This problem can be solved as the 4TP problem in the complex plane. 
Because of importance of the oscillator potential in many 
applications and with the purpose of further development of the WKB 
method, in Ref. \cite{SeQ} the problem has been solved by two 
methods, on the real axis as 2TP problem and then in the complex 
plane as 4TP problem.

Consider the WKB$_0$ solution of the problem.  In the complex plane, 
the problem has two cuts, between turning points $r_1$, $r_2$ and 
$r_3$, $r_4$. To apply residue theory for the phase space integral we 
need to take into account all zeroes of the {\em wf} in the complex 
plane, i.e. the contour $C$ has to include both cuts. The 
quantization condition (\ref{genC}) has the form

\begin{equation}  \label{Io} I =\oint_C\sqrt{2mE - (m\omega r)^2 - 
\frac{\vec M^2}{r^2}}dr = 2\pi\hbar\left(N+\frac\mu 4\right), 
\end{equation} 
where $\mu=4$ and $C$ is the contour about the cuts at $r<0$ and 
$r>0$, respectively. The number $N=n_{r<0}+n_{r>0}$, where $n_{r<0}$ 
and $n_{r>0}$ are the numbers of zeroes of the {\em wf} at $r<0$ and  
$r>0$, respectively.  For the 3D harmonic oscillator, because of 
symmetricity of the potential, we have $n_{r<0}=n_{r>0}=n_r$, i.e. 
the total number of zeroes is $N=2n_r$.

Integral (\ref{Io}) is reduced to the above case of the Coulomb 
potential with the help of the replacement $z=r^2$. Integration 
result is $I=\pi(E/\omega -M)/2$ and we obtain, for the energy 
eigenvalues,

\begin{equation}  \label{Eosc}    
E_n = \omega\left[2\hbar\left(n_r + \frac 12\right) + M\right].
\end{equation}
So far, as $M = (l+1/2)\hbar$, we get the exact energy spectrum for 
the isotropic oscillator. Energy of the ground state is $E_0 
=\omega(\hbar +M_0)$, where $M_0$ is the contribution of quantum 
fluctuations of the angular momentum \cite{SeF}. The asymptotic 
eigenfunctions have the form (\ref{wflo}), where the eigenmomentum 
$P_n$ is calculated with the use of the energy eigenvalues 
(\ref{Eosc}).

\subsection{\label{Hult} The Hulth\'en potential
$V(r)=-V_0e^{-r/r_0}/(1-e^{-r/r_0})$}

The Hulth\'en potential is of a special interest in atomic and 
molecular physics. The potential is known as an ``insoluble'' by the 
standard WKB method potentials, unless one supplements it with 
Langer-like corrections. The radial problem for this potential is 
usually considered at $l=0$. However, in the approach under 
consideration, the quasiclassical method results in the nonzero 
centrifugal term at $l=0$ and allows to obtain the analytic result 
for all $l$.

The quantization condition (\ref{genC}) for the Hulth\'en potential 
is

\begin{equation}  \label{Ihul}
I=\oint\sqrt{2m\left( E + V_0\frac{e^{-r/r_0}}
{1-e^{-r/r_0}}\right) -\frac{\vec M^2}{r^2}}dr =  \nonumber  \\
2\pi \hbar\left(n_r+\frac 12\right).~
\end{equation}
In the region $r>0$, this problem has two turning points $r_1$ and 
$r_2$. The phase-space integral (\ref{Ihul}) is calculated 
analogously to the above case. Introducing the new variable $\rho = 
r/r_0$, we calculate the contour integral in the complex plane, where 
the contour $C$ encloses the classical turning points $\rho_1$ and 
$\rho_2$. Using the method of stereographic projection, we should 
exclude the infinities at $r=0$ and $\infty$ outside the contour $C$.  
Excluding these infinities we have, for the integral (\ref{Ihul}), 
$I=I_0+I_\infty$, where $I_0=-2\pi M$ and $I_\infty$ is calculated 
with the help of the replacement $z=e^\rho -1$ \cite{SeSe},

\begin{eqnarray}  \label{IhulClo}
I = \oint \sqrt{2mr_0^2\left(E + V_0
\frac{e^{-\rho }}{1-e^{-\rho }}\right) - \frac{\vec M^2}
{\rho ^2}}d\rho = \nonumber  \\
-2\pi M + 2\pi r_0\sqrt{-2m}\left[ -\sqrt{-E} + \sqrt{-E+V_0}\right].~
\end{eqnarray}

Substituting the integration result into Eq. (\ref{Ihul}), we 
immediately get the exact energy spectrum

\begin{equation}   \label{Ehul}
E_n=-\frac 1{8mr_0^2}\left(\frac{2mV_0r_0^2}N - N\right)^2.
\end{equation}
where $N=(n_r+1/2)\hbar +M$ is the principal quantum number.  Setting 
in (\ref{Ehul}) $M=0$, we arrive at the energy eigenvalues obtained 
from known exact solution of the Schr\"odinger's equation at $l=0$. 
However, in our case $M_{min}\equiv M_0 = \hbar/2$ at $l=0$ and the 
principal quantum number is $N = (n_r+1/2)\hbar + M_0$.  As in the 
previous examples, this apparently shows the contribution of the 
quantum fluctuations of the angular momentum into the energy of the 
ground state, $E_0$.

\subsection{\label{Morse} The Morse potential
$V(r)=V_0[e^{-2\alpha (r/r_0-1)}- 2e^{-\alpha (r/r_0-1)}]$}

The Morse potential is usually considered as one-dimensional problem 
at $l=0$. In this case the problem has two turning points (note that 
the left turning point, $r_1$, is negative) and can be solved 
exactly.  In the general case, for $l>0$, we have an ``insoluble'' 
$4$TP problem.

For this potential, let us consider, first, the radial Schr\"odinger 
equation, which does not contain the centrifugal term at $l=0$,

\begin{equation}   \label{shMor}
\left(-i\hbar\frac d{dr}\right)^2R(r) = \nonumber  \\
2m\left[E-V_0 e^{-2\alpha(r-r_0)/r_0}+2V_0e^{-\alpha (r-r_0)/r_0}
\right]R(r).~
\end{equation}
The quantization condition (\ref{qc2}) appropriate to this equation 
is

\begin{equation}  \label{IMor}
\oint_C\sqrt{2m[E - V_0e^{-2\alpha (r-r_0)/r_0} +
2V_0e^{-\alpha (r-r_0)/r_0}]}dr =  \nonumber  \\
2\pi\hbar\left(n_r+\frac 12\right).~
\end{equation}
Introducing a variable $x=e^{-\alpha (r-r_0)/r_0}$, we reduce the 
phase-space integral to the well known one. Sequential simple 
calculations result in the exact energy eigenvalues

\begin{equation}   \label{EMor}
E_n = -V_0\left[1-\frac{\alpha\hbar(n_r +\frac 12)}
{r_0\sqrt{2mV_0}} \right]^2.
\end{equation}

Now, let us deal with Eq. (\ref{RadEq}) for this potential, which 
contains the non-vanishing centrifugal term, $\hbar^2/4r^2$, at 
$l=0$. In this case we have an ``insoluble'' $4$TP problem. In the 
complex plane, the problem has two cuts ($\nu=2$), at $r<0$ and 
$r>0$.  Therefore, we apply the $4$TP quantization condition 
(\ref{genC}),

\begin{equation}  \label{IMoM}
\oint \sqrt{2mr_0^2[E-V_0e^{-2\alpha (\rho-1)}+2V_0e^{-\alpha (\rho -1)}]
-\frac{\lambda ^2}{\rho ^2}}d\rho =  \nonumber \\
4\pi\hbar \left(n_r +\frac 12\right)=I,~~~
\end{equation}
where we have introduced the new variable, $\rho =r/r_0$. The contour 
$C$ encloses the two cuts, but does not enclose the point $r=0$. To 
calculate the integral (\ref{IMoM}) we use the method of 
stereographic projection. For this, we should exclude the 
singularities outside the contour $C$, i.e. at $r=0$ and $\infty$. 
Excluding these infinities we have, for the integral (\ref{IMoM}) 
\cite{SeSe},

\begin{equation}   \label{IMocl}
I = -2\pi M -\frac{2\pi r_0}{\alpha}\left(\sqrt{-2mE} -
\sqrt{2mV_0}\right),
\end{equation}
and for the energy eigenvalues this gives

\begin{equation} \label{EMorM}
E_n = -V_0\left[1-\alpha\frac{2\hbar(n_r+\frac 12) +
M}{r_0\sqrt{2mV_0}} \right]^2.
\end{equation}

Setting in (\ref{EMorM}) $l=0$, we obtain,

\begin{equation}  \label{EMoM0}
E_n = -V_0\left[1-\frac{\alpha[2\hbar(n_r+\frac 12)+M_0}
{r_0\sqrt{2mV_0}} \right]^2.
\end{equation}
Equation (\ref{EMoM0}) for $E_n$ is different from the expression 
(\ref{EMor}) obtained from solution of radial Schr\"odinger equation 
for the Morse potential at $l=0$. This difference is caused by the 
nonzero centrifugal term $\hbar^2/4r^2$ in the radial equation 
(\ref{RadEq}) at $l=0$. Thus we obtain two results for the Morse 
potential: the known exact eigenvalues (\ref{EMor}) obtained from 
solution of radial equation at $l=0$ and another result (\ref{EMorM}) 
obtained from solution of Eq.  (\ref{RadEq}) for all $l$.

\subsection{\label{RelCorn} The relativistic Cornell problem,
$V(r)= -\tilde\alpha/r +\kappa r$}

In this paragraph, we use the WKB$_0$ approach to reproduce the exact 
energy spectrum for the famous funnel type potential (Cornell 
potential) \cite{Eich} known as one which is ``insoluble'' exactly in 
terms of special functions. This potential is one of a special 
interest in high energy hadron physics, quarkonium physics, and quark 
potential models. Its parameters are directly related to basic 
physical quantities of hadrons: the universal Regge slope $\alpha 
'\simeq 0.9\,($GeV$/c)^{-2}$ of light flavours and one-gluon-exchange 
coupling strength $\alpha_s$ for heavy quarkonia. A closer 
inspections reveals \cite{LuS} that all phenomenologically acceptable 
"QCD-inspired" potentials are only variations around the funnel 
potential.

In relativistic theory, the Cornell potential represents the 
so-called ``insoluble'' 4TP problem. It is insoluble also from the 
viewpoint of the conventional 2TP WKB approximation. Show that the 
WKB$_0$ method and the quantization condition (\ref{genC}) derived 
above give the asymptote of the exact solution and yield the exact 
energy spectrum for the potential. To solve the problem, we use the 
relativistic semiclassical wave equation obtained in Ref. 
\cite{SeRe}.  For a two-particle system of equal masses, $m_1=m_2=m$, 
interacting by means of the Cornell potential, the relativistic 
radial semiclassical wave equation is ($\hbar=c=1$)

\begin{equation}   \label{qcPot}
\left(-i\frac d{dr}\right)^2\tilde R(r) =  \nonumber  \\
\left[\frac{E^2}4 - \left(m -\frac{\tilde\alpha}r +\kappa r\right)^2
-\frac{(l+\frac 12)^2}{r^2}\right]\tilde R(r),
\end{equation}
where $\tilde\alpha =4\alpha_s/3$. The quantization condition 
(\ref{genC}) appropriate to Eq. (\ref{qcPot}) is

\begin{equation}  \label{Int}
I =\oint_C\sqrt{\frac{E^2}4 -\left(m -\frac{\tilde\alpha}r+\kappa
r\right)^2 - \frac{(l+\frac 12)^2}{r^2}}dr = \nonumber  \\
4\pi\left(n_r+\frac 12\right).
\end{equation}

As in above examples, to calculate the phase-space integral 
(\ref{Int}) in the complex plane, we use the method of stereographic 
projection.  Chose a contour $C$ enclosing the cuts (and, therefore, 
zeros of the {\em wf}) at $r>0$ and $r<0$ between the turning points 
$r_1$, $r_2$ and $r_3$, $r_4$, respectively. Outside the contour $C$, 
the problem has two singularities, i.e. at $r=0$ and $\infty$. Using 
the method of stereographic projection, we have, for the integral 
(\ref{Int}), $I=I_0+I_{\infty}$, where

\begin{equation}  \label{I0}
I_0 = -2\pi\sqrt{(l + 1/2)^2 + \tilde\alpha^2}
\end{equation}
and the integral $I_{\infty}$ is calculated in closed form with the 
help of the replacement of variable, $z=1/r$, that gives

\begin{equation}  \label{Iinf}
I_{\infty} = 2\pi\left(\frac{E^2}{8\kappa} + \tilde\alpha\right).
\end{equation}
Therefore, for $E_n^2$, we obtain

\begin{equation}  \label{EnExac}
E_n^2 = 8\kappa\left[2\left(n_r+\frac 12\right) +
\sqrt{\left(l+\frac 12\right)^2 + \tilde\alpha^2}-\tilde\alpha\right],
\end{equation}
which is the exact result for the Cornell potential.

It is an experimental fact that the dependence $E_n^2(l)$ is linear 
for light mesons (linear Regge trajectories \cite{Se1}). However, at 
present, the best way to reproduce the experimental masses of 
particles is to rescale the entire spectrum assuming that the masses 
$M_n$ of the mesons are expressed by the relation $M_n^2=E_n^2-C^2$, 
where $E_n^2=8\kappa(2n_r+l+3/2)$ is the exact energy eigenvalues for 
the linear potential and $C$ is a constant energy (additional free 
parameter) \cite{SC}. The equation for $M_n^2$ is used to shift the 
spectra and appears as a means to simulate the effects of unknown 
structure approximately.

At the same time, the required shift of the spectra naturally follows 
from Eq. (\ref{EnExac}). The Coulomb-like term, $-\tilde\alpha/r$ of 
the Cornell potential takes into account the week coupling effect 
that allows to describe the Regge trajectories observed at 
experiment. For light mesons one may expect that the $q\bar q$ bound 
states will feel only the linear part of the potential, which gives 
the main contribution to the binding energy. We thus assume that the 
Coulomb term can be considered as a small perturbation. Then we 
obtain, from Eq. (\ref{EnExac}),

\begin{equation}  \label{EnLin}
E_n^2 = 8\kappa\left(2n_r +l -\frac 43\alpha_s +\frac 32\right).
\end{equation}

Equation (\ref{EnLin}) does not require any additional free parameter 
and reproduces the linear Regge trajectories as it observed at 
experiment. We obtain the necessary shift with the help of the term 
$-32\kappa\alpha_s/3$ which is the result of interference of the 
Coulomb and linear terms of the interquark potential. Note that we 
obtain the correct sign (minus) for this term only in the case of the 
{\em scalar-like} potential. Equation (\ref{EnExac}) describes the 
light meson trajectories with the accuracy $\simeq 1\%$ 
\cite{Se1,Se2} (see Table 1).

\medskip \begin{center} \medskip
\centerline{Table 1. $\rho$ family mesons} \medskip
\begin{tabular}{lll}
\hline
\ \ \ ${\rm State}$ & ${\rm E_n}^{theor}$ & ${\rm E_n}^{exp}$\\
\hline
\hline
$\ \ \ 1^3S_1$ & \ \ $0.763$ & $0.768^*$ \\
$\ \ \ 1^3P_2$ & \ \ $1.319$ & $1.318^*$ \\
$\ \ \ 1^3D_3$ & \ \ $1.703$ & $1.691^*$ \\
$\ \ \ 1^3F_4$ & \ \ $2.014$ & $2.037  $ \\
$\ \ \ 1^3G_5$ & \ \ $2.284$ & $\ \ -  $ \\
$\ \ \ 2^3S_1$ & \ \ $1.703$ & $1.700^*$ \\
$\ \ \ 2^3P_2$ & \ \ $2.014$ & $\ \ -  $ \\
$\ \ \ 2^3D_3$ & \ \ $2.284$ & $\ \ -  $ \\
$\ \ \ 2^3F_4$ & \ \ $2.525$ & $\ \ -  $ \\
\hline
\end{tabular} \end{center}
The symbol $^{*}$ denotes the masses of the fitted states; parameters 
of the fit are: $\alpha_s=0.75\pm 0.03$ and $\kappa=(0.14\pm 0.01)$ 
GeV$^2$.

\section{\label{Concl} Conclusion}

We have considered solution of the Schr\"odinger's equation in the 
zero order WKB approximation. One might treat the WKB$_0$ solution as 
a very crude quasiclassical approximation. However, the WKB$_0$ 
solution represents the main term of the asymptotic series in theory 
of the second-order differential equations. A proper treatment of the 
WKB$_0$ approximation opens the way to a simple asymptotic solution 
of the wave equation.

It is a well known fact that the exact eigenvalues can be defined 
with the help of the asymptotic solution. The WKB$_0$ approach gives 
a simplest way to find the asymptote of the exact solution directly, 
without solution of the Schr\"odinger's equation, and results in 
quantization of the classical action.

We have observed several important advantages of the zeroth 
approximation with reference to the commonly used first-order WKB 
approximation. Because the general WKB$_0$ solution (\ref{psi0}) has 
no divergencies at the classical turning points, this allows a 
considerably simpler derivation of connection formulas and 
quantization condition. Using the general requirements of continuity 
and finiteness for WKB$_0$ function $\psi_0(x)$ and its derivative 
$\psi_0^\prime(x)$, we have derived a simple connection formulas, 
which allowed us to build the WKB$_0$ {\it wf} in the whole region 
and the corresponding quantization condition for the classical 
action.

If WKB$_1$ approximation is applicable at a distance from the turning 
points satisfying the condition $\left|x-x_0\right| \gg\lambda/4\pi$, 
the WKB$_0$ solution considered in this work is applicable in the 
whole region.  To demonstrate advantages of the WKB$_0$ approximation 
we have derived not only well known 2TP but, also, 
multi-turning-point quantization condition.  This approach has been 
also used to derive the quantization conditions for periodic 
potentials \cite{SeSu} and can be easily generalized for nonseparable 
problems.

The most important advantage of the WKB$_0$ approach is that the 
WKB$_0$ quantization condition successfully reproduces the exact 
energy spectra for {\it all} known solvable potentials and 
``insoluble'' potentials, too. To demonstrate efficiency of the 
WKB$_0$ method, we have solved several classic problems. The method 
successfully reproduces the exact energy spectrum not only for 
solvable one-dimensional and spherically symmetric potentials but, 
also, for ``insoluble'' potentials with more than two turning points.

For discrete spectrum, the WKB$_0$ eigenfunctions have written in 
terms of elementary functions in the form of a standing wave, 
$A_n\cos(k_nx+\delta_n)$. The fact the eigenfunctions can be written 
in such form means the solution describes free motion of a 
particle-wave in the enclosure. This, in turn, means that the 
eigenmomentum is the constant of motion and the corresponding 
coordinate is cyclical.

The zeroth approximation reproduces all basic results of quantum 
mechanics:  eigenvalues, asymptotes, probabilities, etc.  Obviously, 
this is more than just an approximation. This is, at least, an 
alternative quantization method. What is more, we do not even need to 
solve the Schr\"odinger's equation to find the asymptote. After we 
have fixed the classical action, using the existence/quantization 
formulas, we can write down the quantization condition for an 
arbitrary potential.

This approach can be easily generalized for the non-separable 
problems. In this case the classical action can not be written as sum 
of actions for each variable (degree of freedom). As a result, we 
cannot have a unique quantum number for each degree of freedom.  We 
will have one quantum number for the non-separable variables, and 
unique quantum number for each separable variable.  Asymptote will 
have the same form, but the existence formulas (quantization 
condition) will be multi-dimensional integral over the non-separable 
variables.

In the WKB$_0$ method, we use the same technique for all types of 
problems. The same simple rules formulated for 2TP problems work for 
$\mu$TP problems, as well. In this sense, the WKB$_0$ approach can be 
considered as a general method for finding the asymptote of the 
Schr\"odinger's equation.

{\it Acknowledgments}.  The author thanks Prof. Uday P. Sukhatme for 
kind invitation to visit the University of Illinois at Chicago where 
a part of the work has been done and, also, for useful discussions 
and valuable comments. I should also like to thank Prof. A.A. Bogush 
for support and constant interest to this work.

This work was supported in part by the Belarusian Fund for 
Fundamental Research.

\end{document}